\documentclass[amsmath,amssymb,floats,twocolumn]{revtex4}

\usepackage{graphicx}
\usepackage{amssymb}
\usepackage{amsmath}
\usepackage{hyperref}
\usepackage{footnote}
\usepackage{color}
\usepackage{multirow}
\pdfoutput=1
\begin{document}

\title{Improved superconducting qubit coherence with high-temperature substrate annealing}
\author{Archana Kamal$^{1}$}
\email{akamal@mit.edu}
\author{Jonilyn L. Yoder$^{2}$}
\author{Fei Yan$^{1}$}
\author{Theodore J. Gudmundsen$^{2}$}
\author{David Hover$^{2}$}
\author{Adam P. Sears$^{2}$}
\author{Paul Welander$^{2}$}
\altaffiliation{Current address: SLAC National Accelerator Laboratory, 2575 Sand Hill Road, Menlo Park, CA 94025, USA}
\author{Terry P. Orlando$^{1}$}
\author{Simon Gustavsson$^{1}$}
\author{William D. Oliver$^{1,2}$}
\affiliation{
$^{1}$Research Laboratory of Electronics, Massachusetts Institute of Technology, Cambridge, MA 02139, USA\\
$^{2}$ MIT Lincoln Laboratory, 244 Wood Street, Lexington, MA 02420, USA}
\begin{abstract}
We assess independently the impact of high-temperature substrate annealing and metal deposition conditions on the coherence of transmon qubits in the standard 2D circuit-QED architecture. We restrict our study to devices made with aluminum interdigital capacitors on sapphire substrates. We record more than an order-of-magnitude improvement in the relaxation time of devices made with an annealed substrate, independent of whether a conventional evaporator or molecular beam epitaxy chamber was used to deposit the aluminum. We also infer similar levels of flux noise and photon shot noise through dephasing measurements on these devices. Our results indicate that substrate annealing plays a primary role in fabricating low-loss qubits, consistent with the hypothesis that substrate-air and substrate-metal interfaces are essential factors limiting the qubit lifetimes in superconducting circuits.
\end{abstract}
\maketitle

Significant advances in the coherence of superconducting qubits~\cite{OliverMRS2013} have facilitated their emergence as a serious candidate for quantum information processing applications~\cite{DevoretSchoelkopf2013}. These advances have followed in part from extensive studies highlighting the role of device geometry~\cite{Wenner2011,Geerlings2012,BarendsXmon}, the electromagnetic environment~\cite{Paik2011,Wang2015}, and constituent materials~\cite{Megrant2012,Chang2013,Steele2014,UCSBReport,Tournet2016} on energy loss in superconducting circuits. In particular, superconducting coplanar waveguide resonators and qubits with planar geometries have exhibited substantially improved performance when fabricated using molecular beam epitaxy (MBE)-grown aluminum, deposited on sapphire substrates annealed at high temperatures~\cite{Megrant2012,BarendsXmon}. While this improvement is likely due to a reduction in the surface loss at the various metal and substrate interfaces~\cite{Wenner2011}, it remains an open question whether the reduction arises primarily from substrate preparation, metal growth conditions, or their combination.
\par
In this Letter, we study independently the effects of substrate preparation and metal growth conditions to identify which (if either) is the dominant contributor to observed improvements in qubit coherence. To do this, we use frequency-tunable transmons shunted with an interdigital capacitor (IDC) as a benchmark qubit~\cite{Koch2007}.
We first perform an {\em annealing test \em}:
a $T_1$ characterization of transmons fabricated on annealed and non-annealed sapphire substrates using the same conventional evaporator to deposit the aluminum IDCs.
We then perform a {\em deposition test \em}:
a $T_1$ characterization of transmons with IDCs comprising aluminum deposited in higher-vacuum (MBE) and lower-vacuum (conventional evaporator) chambers on similarly annealed sapphire substrates.
Additionally, we perform a {\em geometry test \em} by comparing the $T_1$ of transmons with different IDC finger sizes. To make a comparison across devices at different frequencies, we parameterize $T_1$ as a frequency-independent quality factor $Q$.
We obtain $Q>10^{6}$ for qubits on annealed substrates \emph{irrespective} of the deposition method, whereas the $Q$ is more than an order of magnitude lower for devices fabricated on non-annealed substrates.
We furthermore demonstrate that $Q$ increases with larger IDC size, consistent with earlier works~\cite{Wenner2011,Geerlings2012,BarendsXmon}.
For completeness, we also measured $T_2$ and found the observed dephasing times to be consistently explained by photon number fluctuations in the resonator, independent of the deposition method and device geometry.
\par
The fabrication steps for the substrate annealing were similar, and subsequent junction patterning was nominally identical for all transmon samples presented in this work. All wafers were cleaned in solvents prior to metal deposition. Then, for the deposition test on annealed wafers, we compared the following approaches:
\begin{itemize}
\item High-vacuum electron-beam aluminum deposition (conventional evaporator): 50-mm sapphire wafers were annealed in a furnace tube with a dry atmosphere environment (80:20 nitrogen:oxygen) at 1100~$^{\circ}$C. After annealing, the wafers were transferred to a standard evaporation chamber (base pressure $10^{-7}$ Torr), where 150 nm of electron-beam-evaporated aluminum was deposited at a rate of $5\;  {\buildrel_{\circ} \over {\mathrm{A}}}/{\rm s}$.

\item Ultra-high vacuum (UHV) aluminum deposition (MBE chamber): 50-mm sapphire wafers were annealed in the MBE growth chamber (base pressure $10^{-11}$ Torr) at 900~$^{\circ}$C. Without breaking vacuum, 235 nm of aluminum was deposited from an effusion cell at a rate of $0.25 \; {\buildrel_{\circ} \over {\mathrm{A}}}/{\rm s}$.  During deposition, the wafer was heated to 150~$^{\circ}$C.
\end{itemize}
In the context of this work, we believe (and our results suggest) that above differences in annealing procedures, attributable to our standard growth practices, do not significantly impact our conclusions. For the annealing test, we compare against a non-annealed substrate with 180 nm of aluminum deposited, concurrently with the Josephson junction fabrication, using double-angle evaporation in the same conventional evaporator. While this comparison admittedly represents a substantial difference in capacitor formation and thus lacks strict experimental control, it does reflect a common fabrication approach used in the community and is therefore of practical interest.
\begin{table}[t!]
\begin{tabular}{cccccc|ccc}
\hline
Sample & Dep- & Anneal & W/S  & $f_{ge}$ & $T_{1}^{\rm meas}$ & $T_{1}^{\rm Purcell}$ & $T_{1}^{\rm diel}$ & $Q^{\rm diel}$\\
 \# & Tool &  & ($\mu$m) & (GHz) &  ($\mu$s) & ($\mu$s) & ($\mu$s) & $\times 10^{6}$\\
\hline
\hline
1 & E & no & 10/10 &  6.5 & {\bf 0.29} & 5 & 0.24 & 0.01\\
2 & E & no & 10/10 &  8.0 & {\bf 0.25} & 1 & 0.30 & 0.01\\
3 & E & yes & 10/10 & 3.68 & 7.2 & 145 & 8 & 0.18\\
4 & M & yes & 10/10 & 5.5 & {\bf 8.8} & 160 & 9 & 0.25\\
& & & & 4.0 & 7.2 & 360 & 7\\
5 & E & yes & 10/20 & 6.3 &  {\bf 14.5} & 24 & 26 & 1.01\\
& & & & 5.59 & 15.7 & 42 & 29 &\\
& & & & 4.16 & 20.2 & 92 & 39 &\\
6 & M & yes & 10/20 & 5.29 &  {\bf 19.7} & 30 & 38 & 1.25\\
& & & & 5.07  & 15.9 & 34 & 39\\
& & & & 4.41 & 18.4 & 48 & 45\\
7 & E & yes & 25/25 & 6.43 & {\bf 13.2} & 24 & 37 & 1.5\\
& & & & 5.01 & 28.7 & 66 & 48\\
& & & & 3.80 & 35.3 & 120 & 63\\
\hline
\end{tabular}
\caption{{\bf $T_{1}$ measurements on 2D transmons: deposition, annealing, and geometry tests.} Samples are numbered 1-7, with the type of deposition chamber, ``E'' (conventional evaporator, lower vacuum) and ``M'' (MBE, higher vacuum), and annealing (yes/no) indicated separately. $W$ and $S$ denote IDC finger width and spacing [see inset of Fig. \ref{FigT1}(a)]; $f_{ge}$ denotes the transmon ground-excited state transition frequency; and $T_{1}^{\rm meas}$ is the measured decay time. Bold values were measured at the respective flux-insensitive spots. Estimates for the Purcell-limited decay time ($T_{1}^{\rm Purcell}$), dielectric-loss-limited decay time ($T_{1}^{\rm diel}$), and the average dielectric quality factor  $Q^{\rm diel}$ are calculated or inferred values (see text). Additional qubit and readout resonator parameters are available in the Supplementary Materials~\cite{Supplement}}
\label{T1table}
\end{table}

\par
%

Representative relaxation times for transmon samples with different types of deposition, annealing, and IDC geometry are shown in Table~\ref{T1table}. To compare qubits at different frequencies, we need to infer a frequency-independent quality factor for each sample. We therefore assume dielectric loss and model it with an equivalent noise admittance seen by the junction~\cite{Supplement}. Assuming the coupling Hamiltonian between the qubit and the dissipative admittance of the form $H = (\Phi_{0}/2\pi) \hat{\varphi}\cdot\hat{I}_{N}$, we calculate the decay rate between desired qubit levels using Fermi's golden rule. Here $\hat{\varphi}$ is the phase operator associated with the Josephson junction and $\hat{I}_{N}$ is the fluctuating current sourced by the admittance. For a transmon circuit with charging energy $E_{C}$ and Josephson energy $E_{J}$ with $E_{J}/E_{C} \gg 1$, the matrix elements between successive levels of the transmon, $\langle j +1 |\varphi | j \rangle= \sqrt{(j+1)/2} (E_{C}/E_{J} (\Phi))^{1/4}$, are those of a simple harmonic oscillator to leading order in $(E_{C}/E_{J})$. This gives a simplified expression for qubit decay rate as~\cite{Supplement}
\begin{eqnarray}
	\Gamma_{g \rightarrow e}^{\rm cap} = p_{S} \frac{\omega_{ge}}{Q_{S}} + p_{J} \frac{\omega_{ge}}{Q_{J}},
	\label{EqGamma}
\end{eqnarray}
with $\omega_{ge} = 2 \pi f_{ge}$. Here ${(Q_{S}, Q_{J})}$ and $p_{S} = C_{S}/(C_{S} + C_{J})$, $p_{J} = C_{J}/(C_{S} + C_{J})$ represent the frequency-independent quality factors and participation ratios associated with the external shunt capacitance $C_{S}$ and the junction self-capacitance $C_{J}$, respectively. Since $C_{S} \ll C_{J}$, we have $p_{J} \ll p_{S} \approx 1$; this implies the qubit decay is dominated by the loss in the shunt capacitor \footnote{Also, the quality factor associated with the junction dielectric, $Q_{J} \geq 10^{7}$, based on results reported in Refs. \cite{Paik2011} and \cite{Kim2011}}. Assuming the capacitive loss to be contributed entirely by the dielectric substrate, we infer a $ Q^{\rm diel} \approx Q_{S}$ as a fit parameter from $T_{1}$ data.
\begin{figure}[t!]
\centering
  \includegraphics[width=\columnwidth]{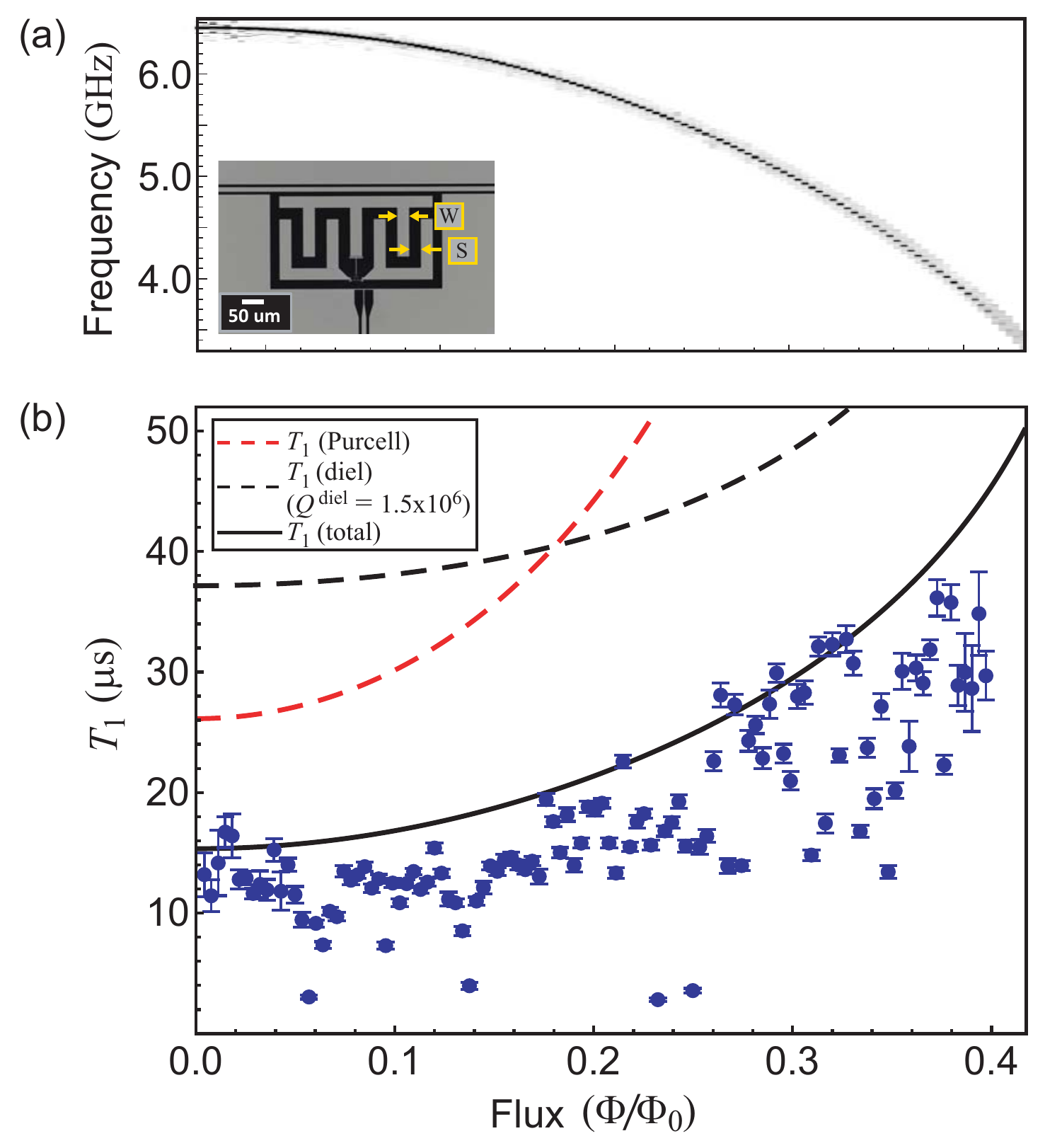}\\
  \caption{{\bf $T_{1}$ spectroscopy for sample 7.} (a) Spectroscopy of $f_{ge}$ vs flux bias of the SQUID loop. The inset is an image of a typical transmon sample showing the IDC geometry. (b) The blue dots show the $T_{1}$ data recorded as a function of qubit frequency. The curves show theoretical predictions for Purcell-limited $T_{1}$ (dashed red), $T_{1}$ limited by dielectric loss in the shunt capacitor (dashed black), assuming a given loss tangent for the dielectric $(Q^{\rm diel})^{-1} = \tan \delta$, and total $T_{1}$ (solid black) obtained by combining the two contributions shown with dashed curves, i.e., $T_{1}^{-1}({\rm total}) = T_{1}^{-1} ({\rm Purcell}) + T_{1}^{-1} ({\rm diel})$ \cite{Supplement}. The upward trend in $T_{1}$ with decrease in $f_{ge}$ is well described by the  dielectric-loss dominated decay model of Eq. (\ref{EqGamma}).}
\label{FigT1}
\end{figure}
\par
%
For the \emph{annealing test}, we compare samples 1, 2 and 3 ---  samples with IDCs nominally identical in geometry and deposited using a conventional low-vacuum evaporator. A key fabrication difference was that, unlike samples 1 and 2, the substrate for sample 3 was furnace-annealed at 1100~$^{\circ}$C prior to aluminum deposition. Based on $T_{1}$ times for these samples, we infer more than an order of magnitude higher $Q^{\rm diel}$ for sample 3 as compared to samples 1 and 2; this underscores a significant improvement in qubit coherence correlated with substrate annealing. We note that, however, that the capacitor fabrication processes also differed substantially (single versus double-angle evaporation with oxidation step), and we cannot rule out that some portion of this improvement is related to these differences.
\par
For the \emph{deposition test}, we compare sample 3 with 4 and sample 5 with 6. The sapphire substrates of all these samples received a high-temperature anneal prior to IDC deposition. However, while the subsequent IDC deposition for samples 3 and 5 was performed in a conventional evaporator, that for samples 4 and 6 was done at ultra-high vacuum in an MBE chamber. As can be seen from data in Table~\ref{T1table}, there is no significant difference in $T_{1}$ or inferred $Q^{\rm diel}$ values between samples 3 and 4 or between samples 5 and 6. These results indicate that there is no significant difference in coherence between qubits obtained with higher-vacuum or lower-vacuum e-beam evaporation of the metal, whereas there is a significant difference between qubits with annealed and non-annealed substrates.
%
\par
Finally for the \emph{geometry test}, we compare $Q^{\rm diel}$ values across samples 3 to 7, all of which were fabricated on annealed substrates and had different IDC geometries. We see a consistent improvement in $Q^{\rm diel}$ values with increased spacing between the IDC fingers from 3 to 7.  This trend further corroborates the importance of substrate quality as increased spacing between the interdigitated fingers reduces the surface participation in the shunt capacitor. For samples 5, 6, 7, the three transmons with largest spacing between IDC fingers, we measured $T_{1}$ times in excess of $10 \; \mu$s, corresponding to an average $Q^{\rm diel} > 10^{6}$ over the entire qubit spectrum \footnote{We also obtained consistent $Q$ values, at single-photon energies, for CPW resonators fabricated on the same wafer as the respective transmon qubits \cite{Supplement}}. The highest $T_{1} = 35 \; \mu$s  is measured on sample 7 -- a transmon prepared using standard low-vacuum evaporation on an annealed surface with a $25\; \mu$m spacing between IDC fingers; detailed measurement of $T_{1}$ variation with qubit transition frequency $\omega_{ge}$ for this device is shown in Fig. \ref{FigT1}. \footnote{The dropouts in $T_{1}$ are consistent with the presence of a strongly coupled spurious coaxial mode, as shown by the increase in Rabi frequency for a given drive power at corresponding flux bias values.}
\par
\begin{figure}[t!]
\centering
  \includegraphics[width=\columnwidth]{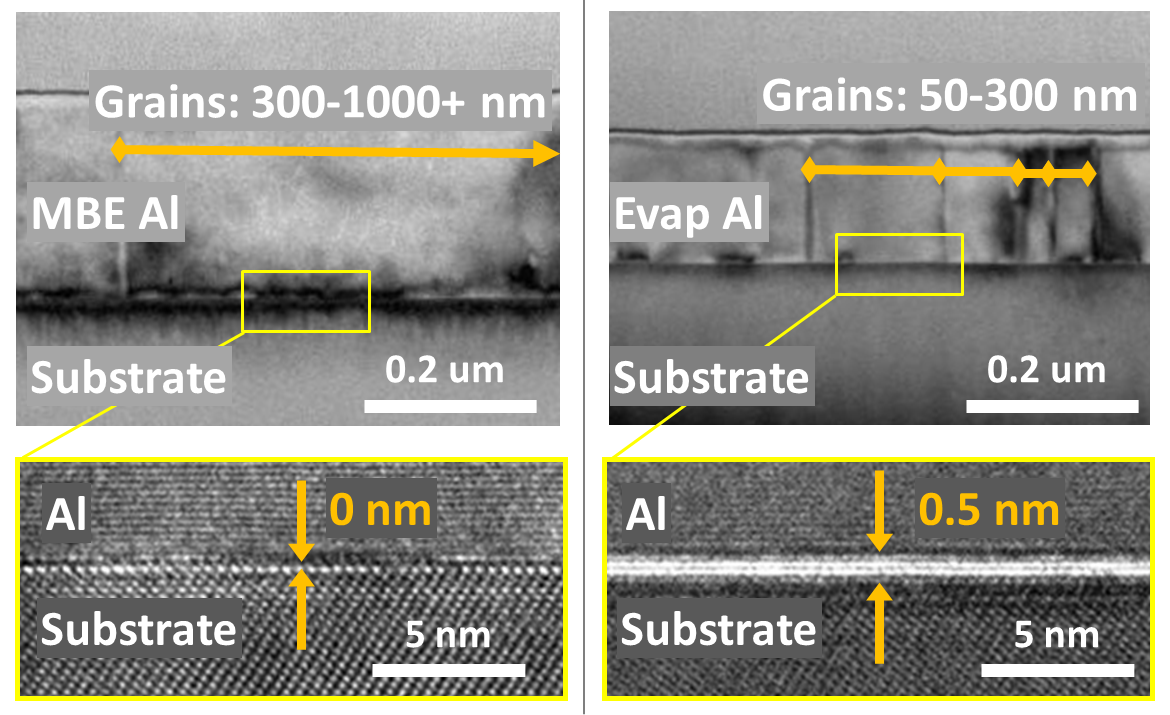}
  \caption{{\bf STEM images of films on annealed sapphire substrates, deposited under high- and low-vacuum conditions.} The films imaged using STEM were cross-sectioned from the capacitor region of sample 6 (left panels) and sample 7 (right panels) respectively. Top panels show the grain structure of the aluminum films. Bottom panels are high-resolution images of the sapphire--aluminum interfaces. Additional STEM cross-sectional images of the aluminum films and interfacial layers are included in the Supplementary Materials \cite{Supplement}.}
\label{FigTEM}
\end{figure}
\par
We further investigated the differences in quality of films and interfaces, between low and high vacuum IDC deposition, using scanning transmission electron microscope (STEM) imaging. Focused ion beam (FIB) lamellae were prepared by cross-sectioning the shunt capacitors of transmons that were concurrently fabricated on the same wafers as samples 6 and 7 in Table \ref{T1table}.  As shown in Fig. \ref{FigTEM}, the aluminum grains for the two films are significantly different; the grains of the film deposited in low-vacuum are approximately 50 nm to 300 nm in width, whereas the aluminum grains for the film deposited in high-vacuum conditions are typically larger than 300 nm reaching up to micron or more in size. High-magnification images of the sapphire-aluminum interfaces of the films detail the presence of a 0.5 nm-thick interfacial layer in sample 7 and no interfacial layer in sample 6. Both of these imaging results are consistent with expectations based on the film annealing and fabrication; the aluminum film in sample 6 was expected to have larger grains since it was deposited at a $10\times$ slower rate than the aluminum film for sample 7. Additionally, since sample 7 was exposed to air between the annealing and evaporation steps, it was expected that a thin contaminant layer would reform at the surface after annealing and lead to an interfacial layer at the metal-substrate interface. The relative purity of this interface (as corroborated by a chemical composition profile of interfaces \cite{Supplement}) may also contribute to the slightly higher $Q^{\rm diel}$ observed in samples 4 and 6, as compared to 3 and 5 with nominally identical geometries; however, this is a relatively minor effect as compared with the presence or absence of substrate annealing.
%
%
%
%
%
%
%
\begin{figure}[t!]
\centering
  \includegraphics[width=\columnwidth]{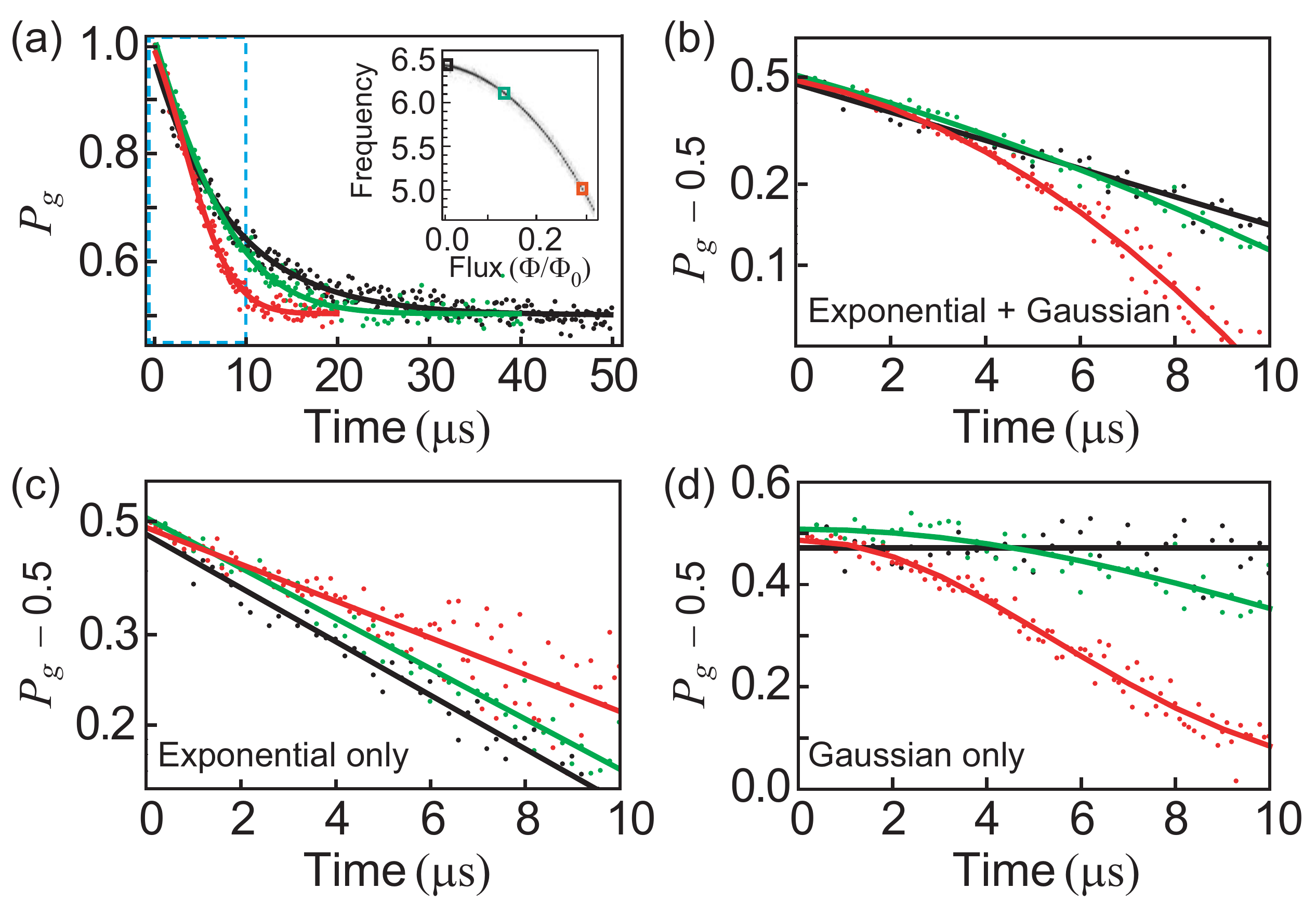}\\
  \caption{{\bf Echo decay measurements for sample 7 at different flux biases.} (a) Echo decay envelopes measured at three different flux biases, indicated with corresponding colored boxes on the spectrum in the inset, along with combined exponential and Gaussian fit functions. (b)-(d) Zoom of the first $10$ $\mu$s record of the traces in (a). Traces with offsets subtracted plotted on a log scale in (b) show the change in shape of the decay envelopes from being purely exponential at the sweet spot (black) to Gaussian at the lower part of the qubit frequency lobe (green and red). Panel (c) [or (d)] shows the data in (b) fitted to purely exponential [or purely Gaussian] decay, post subtraction of the respective Gaussian [or exponential] components from the amplitudes of the echo decay. Agreement of data with solid curves in both (c) and (d), generated using the exponential and Gaussian decay times obtained from the fits in (a), validates the methodology followed to obtain the dephasing times reported in Table \ref{T2table}.}
\label{FigEcho}%
\end{figure}
\begin{table*}[t!]
\begin{tabular}{c|c|c|cc|cc|c|c|c}
\hline
\multirow{2}{*}{Sample} & \multirow{2}{*}{$f_{ge}$ (GHz)} & \multirow{2}{*}{$T_{1}$  ($\mu$s)} & \multicolumn{2}{|c|}{Ramsey} & \multicolumn{2}{|c|}{Echo} & \multirow{2}{*}{$T_{\phi}^{E}$  ($\mu$s)} & \multirow{2}{*}{A $(\mu \Phi_{0})^{2}$} & \multirow{2}{*}{$\langle n \rangle$}\\ \cline{4-7}
& & & $T_{2e}^{R}$ ($\mu$s) & $T_{2g}^{R}$ ($\mu$s) &  $T_{2e}^{E}$ ($\mu$s) & $T_{2g}^{E}$ ($\mu$s) & &\\
\hline
\hline
\multirow{3}{*}{6}  & {\bf 5.29} &  {\bf 19.7} & {\bf 20.9} & {\bf 74.8} & {\bf 23.5} & {\bf 514} & {\bf 58.2} & - &  0.005\\
& 5.07  & 15.9 & 14.5 & 5.7  & 16.9 & 26.8 & 36.1 & 2.0 & 0.009\\
& 4.41 & 18.4 & 20.0 & 2.2 & 18.4 & 10.2 & 36.8 & 2.5 & -\\
\hline
\multirow{4}{*}{7} &  {\bf 6.43} &  {\bf 13.2} & {\bf 4.2} &  {\bf 6.6} & {\bf 8.2} & $ - $ & {\bf 11.8} & - & 0.006\\
& 6.11 & 13.5 & 6.4 & 3.4 & 8.8 & 16.9 & 13.0 & 1.6 & 0.007\\
& 5.01 & 28.7 & 10.7 & 1.5 & 12.5 & 7.5 & 15.9 & 2.5 & -\\
& 3.80 & 35.3 &  $ - $ & $< 1$ & 5.3 & 4.0 & 5.7 & 3.1 & -\\
\hline
\end{tabular}
\caption{{\bf Dephasing measurements for samples 6 and 7.} The symbols $(T_{2e}^{R}, \; T_{2g}^{R})$ denote the exponential and Gaussian decay times for the Ramsey envelope, while  $(T_{2e}^{E}, \; T_{2g}^{E})$ denote the corresponding times for the echo envelopes. The numbers in bold are measured at the flux sweet spots. Also included are respective pure dephasing times $T_{\phi}^{E}$ at each flux bias.}
\label{T2table}
\end{table*}
\par
We also assessed the effect of the low- and high-vacuum metal deposition on qubit dephasing, by performing Ramsey and echo experiments at different flux biases on the highest $Q^{\rm diel}$ devices obtained with each of the protocols --- namely samples 6 and 7 (Table \ref{T2table}). Each of the Ramsey and echo traces was fitted to a combined Gaussian and exponentially decaying envelope of the form $f^{R,E} (t) = e^{-t/ T_{2e}^{R,E}}e^{- \left(t/T_{2g}^{R,E}\right)^{2}}$ to obtain the reported decay times. Figure \ref{FigEcho} details representative data and fits measured at three different flux bias points for sample 7. As the qubit was progressively biased away from the sweet spot, the dephasing times evolved from being white-noise dominated to $1/f$-noise dominated, as reflected by the change in the envelope shapes and the suppression of the Gaussian decay times for both Ramsey and echo traces (Table \ref{T2table}). Assuming a noise spectral density for intrinsic flux fluctuations of the form $S(f) = A/f$, the Gaussian part of the echo decay relates to the total flux noise amplitude $A$ as $(T_{2g}^{E})^{-1} = D_{\Phi} \sqrt{A \log 2}$ \cite{Yoshihara2006} where $D_{\Phi} = |\partial \omega_{ge}/\partial \Phi|$ denotes the slope of the qubit transition with respect to flux. From the $T_{2g}^{E}$ measurements in the flux-noise dominated regime, we infer comparable values of $A = (1.5 \;\mu \Phi_{0})^{2}$ to $(3 \;\mu \Phi_{0})^{2}$ for both samples 6 and 7. These results are consistent with typical flux-noise densities reported previously for superconducting qubits.
\par
Further, the corresponding pure dephasing times, $1/T_{\phi}^{E}= 1/T_{2e}^{E} - 1/(2 T_{1})$, evaluated at/near the flux sweet spot were limited by white noise in both these transmons \footnote{This reflects in the decay of echo envelopes, which is predominantly exponential in nature.}. Under the assumption that this white noise component was due entirely to photon shot noise in the readout resonator, we can infer the (residual) photon population $\langle n \rangle$ in the resonator in the weak dispersive regime ($\chi/\kappa < 1$ for both the samples \cite{Supplement}) as $(T_{\phi}^{E})^{-1} = 8\langle n \rangle \chi^{2}/(\kappa + 4 \chi^{2}/\kappa)$ \cite{Blais2007}. For both samples 6 and 7, we obtain an average photon number $\langle n \rangle \approx 0.007$. Thus, longer dephasing times observed in sample 6 (3-5 times higher than those for sample 7) are consistent with this same level of photon population in the resonator, once we account for the differences in $\kappa$ and $\chi$ for the two samples, i.e., $\kappa_{6}/\kappa_{7} \approx 1.75$ and $\chi_{6}/\chi_{7} \approx 0.5$. These estimates for background shot noise magnitude are in agreement with that obtained from independent and detailed measurements of the shot noise spectrum performed on flux qubits in the same setup \cite{Yan2015}. They correspond to effective temperatures of $\sim 80$ mK for the resonator mode. Similar temperatures and thermal photon populations have been reported previously in other cQED setups \cite{Bertet2005,Yamamoto2014}.
\par
In summary, we studied independently sapphire substrate annealing (annealing versus no annealing) and aluminum metal deposition (high versus low vacuum growth chamber) to identify which (if either) is the dominant contributor to observed improvements in qubit coherence.
We found that annealing the substrate provided a substantial improvement in $T_1$, irrespective of whether the aluminum was grown in a conventional evaporator (low vacuum chamber) or an MBE system (high vacuum chamber).
These results are consistent with the hypothesis that metal-substrate and substrate-air interfaces (rather than the bulk metal) are the dominant loss channels in contemporary superconducting qubits and resonators.
We furthermore demonstrated improvements to $T_1$ by altering the capacitor geometry, and thereby its electric field distribution, to reduce the participation of those loss channels.
In the context of dephasing, we found similar levels of flux noise and resonator photons for qubits on annealed substrates, irrespective of the type of deposition.
We note that these measurements were performed on standard IDC geometries for transmons; based in part on these results, we have successfully employed substrate annealing and optimized capacitor designs to achieve $T_{1} \geq 50 \;\mu s$ and $T_{2e}^{E} \approx 2 T_{1}$ in capacitively shunted flux qubits~\cite{Yan2015}. Our results strongly motivate more systematic studies of annealing on different qubit designs and materials, as well as prospects of high-quality substrate preparation method alternatives to MBE deposition.
\par
The authors wish to acknowledge P. Baldo for fabrication support, G. Fitch for help with design layout, and V. Bolkhovsky and A. J. Kerman for useful discussions. This research was funded by the Office of the Director of National Intelligence (ODNI), Intelligence Advanced Research Projects Activity (IARPA) via MIT Lincoln Laboratory under Air Force Contract No. FA8721-05-C-0002. The views and conclusions contained herein are those of the authors and should not be interpreted as necessarily representing the official policies or endorsements, either expressed or implied, of ODNI, IARPA, or the US Government. The U.S. Government is authorized to reproduce and distribute reprints for Governmental purpose notwithstanding any copyright annotation thereon.

%
%

%
\end{document}


\title{Supplementary material for\\
Improved superconducting qubit coherence with high-temperature substrate annealing}
\author{Archana Kamal$^{1}$}
\email{akamal@mit.edu}
\author{Jonilyn L. Yoder$^{2}$}
\author{Fei Yan$^{1}$}
\author{Theodore J. Gudmundsen$^{2}$}
\author{David Hover$^{2}$}
\author{Adam P. Sears$^{2}$}
\author{Paul Welander$^{2}$}
\altaffiliation{Current address: SLAC National Accelerator Laboratory, 2575 Sand Hill Road, Menlo Park, CA 94025, USA}
\author{Terry P. Orlando$^{1}$}
\author{Simon Gustavsson$^{1}$}
\author{William D. Oliver$^{1,2}$}
\affiliation{
$^{1}$Research Laboratory of Electronics, Massachusetts Institute of Technology, Cambridge, MA 02139, USA\\
$^{2}$ MIT Lincoln Laboratory, 244 Wood Street, Lexington, MA 02420, USA}
\maketitle
%
\section{Transmon parameters}
%
\begin{table}[h!]
\begin{tabular}{c|cccccc}
Sample & $E_{J}$  & $E_{c}$ & $\omega_{ge}^{\rm max}/2\pi$ & $\omega_{\rm res}/2\pi$ &  $\kappa/2\pi$ &  $g/2\pi$ \\
 &  (GHz) &  (GHz) & (GHz)  & (MHz) & (MHz) & (MHz)\\
\hline
\hline
1 & 16.0 & 0.33 & 6.53 & 8520 & 5.02 & 190\\
2 & 26.7 & 0.30 & 8.00 & 8554 & 2.50 & 153\\
3 & 14.5 & 0.31 & 6.44 & 8528 & 1.37 & 167\\
4 & 12.6 & 0.34 & 5.54 & 8542 & 0.87 & 103\\
5 & 14.8 & 0.33 & 6.30 & 8509 & 1.45 & 154\\
6 & 10.8 & 0.32 & 5.29 &  8570 & 3.40 & 130\\
7 & 14.3 & 0.36 & 6.43 & 8537 & 1.95 & 120
\end{tabular}
\caption{{\bf Parameters for the transmon qubits.} All the parameters were measured (or inferred) from qubit and resonator spectroscopies.}
\label{Paramtable}
\end{table}
The transmon $E_{J}, E_{C}$ were inferred from qubit spectroscopy. The dispersive coupling strength $g$ was estimated from qubit-induced Lamb shifts of the respective resonators, which were verified with two-qubit splitting and/or direct measurement of $\chi$.
%
\section{Transmon decay rate: Equivalent Noise circuit model}
%
\subsection{Dielectric loss}
%
\begin{figure}[b!]
\centering
  \includegraphics[width=0.7\columnwidth]{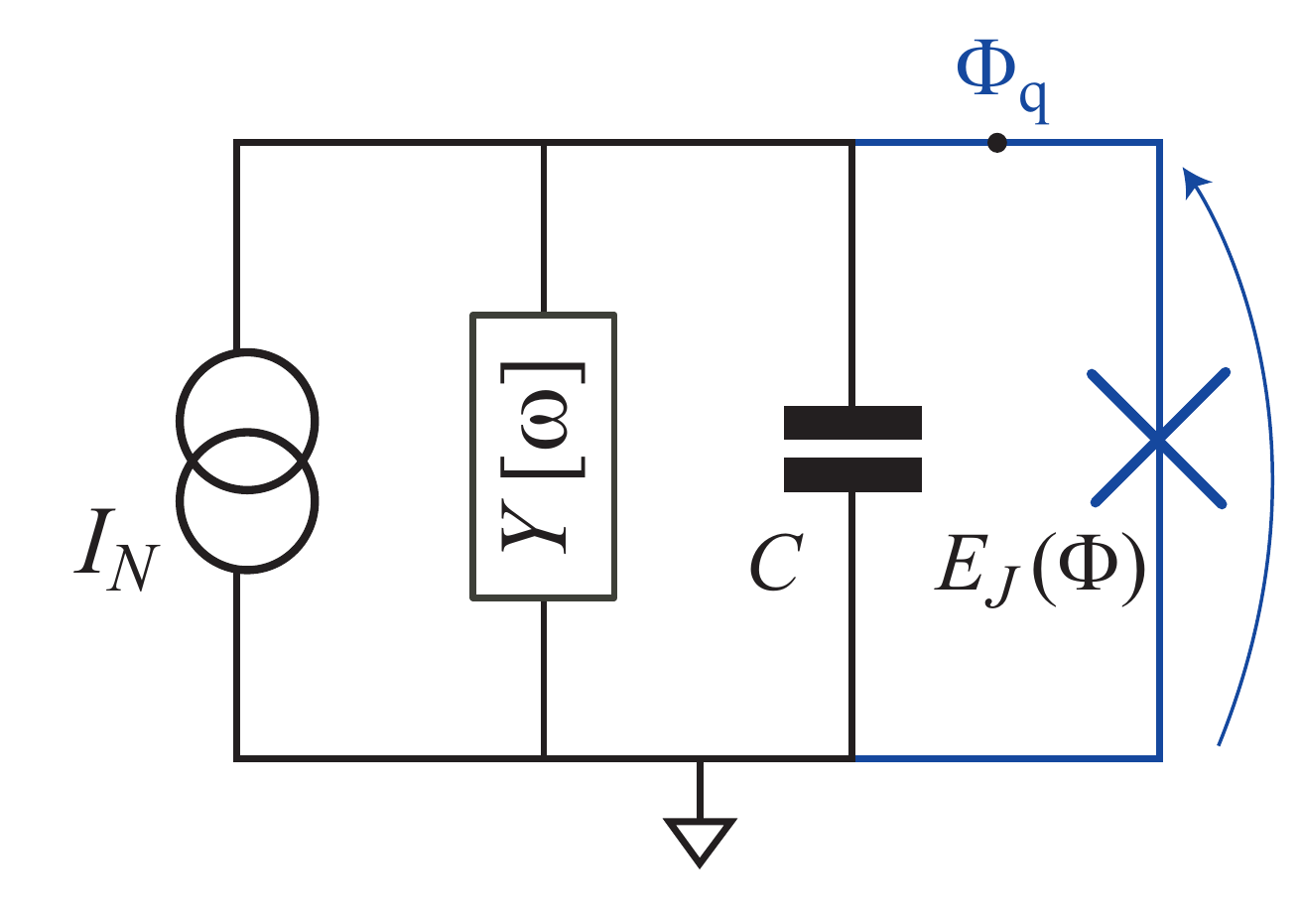}\\
  \caption{{\bf Noise equivalent circuit for a transmon.} The dielectric loss from the capacitor is modelled as a frequency-dependent admittance $Y [\omega]$ seen by the qubit junction (shown in blue). The noise current $I_{N}$ sourced by this admittance couples to the node flux across the junction as described by the Hamiltonian in Eq. (\ref{EqHcoup}).}
  \label{FigNodePhiDiel}
\end{figure}
We consider the noise equivalent circuit model for the qubit as shown in Fig. \ref{FigNodePhiDiel}. This particular choice of representation is motivated by the following reasoning. Consider the total admittance of a parallel-plate capacitive element,
\begin{eqnarray}
	Y(\omega) = j \omega \epsilon (\omega) \frac{A}{d},
\end{eqnarray}
where $A, d$ represent the area and the distance between the plates and $\epsilon(\omega)$ denotes the permittivity associated with the dielectric between the plates. A lossy dielectric can be represented by a complex permittivity of the form $\epsilon(\omega) =\epsilon'(\omega) + j \epsilon'' (\omega)$, which allows us to resolve the total admittance as a combination of a dispersive susceptance,
\begin{eqnarray}
	{\rm Im} \{Y(\omega)\} = \omega \epsilon'(\omega) \frac{A}{d} =\omega C(\omega)
	\label{EqReYCap}
\end{eqnarray}
and a dissipative conductance,
\begin{eqnarray}
	{\rm Re} \{Y(\omega)\} &=& \omega \epsilon''(\omega) \frac{A}{d} 
	= \omega C(\omega) \frac{\epsilon''(\omega)}{\epsilon'(\omega)} \nonumber\\
	&=& \omega C(\omega) \tan \delta(\omega) = \frac{\omega C(\omega)}{Q_{\rm diel} (\omega)}.
	\label{EqReYCap}
\end{eqnarray}
The two branches of the noise admittance in Fig. \ref{FigNodePhiDiel} represent the real and imaginary part of the capacitive admittance seen by the qubit. The coupling between this noise admittance and the qubit can be described by an interaction Hamiltonian of the form,
\begin{eqnarray}
    H_{I} &=& \hat{\Phi}_{q}\cdot \hat{I}_{N} \nonumber\\
    &=& \frac{\Phi_{0}}{2 \pi} \hat{\varphi}\cdot \hat{I}_{N},
\label{EqHcoup}
\end{eqnarray}
where $\varphi$ denotes the dimensionless flux/phase across the junction and $I_{N}$ denotes the amplitude of noise current associated with $Y[\omega]$. Fermi's Golden Rule (FGR) allows us to calculate the decay (or transition) rate between the two levels of the qubit, g, e, due to perturbation by $H_{I}$ as
\begin{eqnarray}
    \Gamma_{g \rightarrow e}^{\rm cap}
    &=&  \frac{1}{\hbar^{2}}\left(\frac{\Phi_{0}}{2 \pi}\right)^{2} |\langle e|\varphi|g\rangle|^{2}
S_{II} [\omega_{ge}].
    \label{EqFGR}
\end{eqnarray}
Here $|\langle e|\varphi|g\rangle|$ denotes the magnitude of the (dimensionless) flux matrix element across the junction calculated between the qubit ground state $|g\rangle$ and excited state $|e\rangle$. The $S_{II} [\omega_{ge}]$ denotes the quantum spectral density of fluctuations associated with the current noise source in parallel with the noise resistance (Fig. \ref{FigNodePhiDiel})
\begin{eqnarray}
    S_{II} [\omega] &=& \hbar \omega {\rm Re}[Y [\omega]] \left[\coth\left(\frac{\hbar \omega}{2 k_{B}T}\right) + 1\right]\\
&\xrightarrow{T \rightarrow 0} & 2 \hbar \omega {\rm Re}[Y [\omega]].
\label{EqNoiseSpec}
\end{eqnarray}
Using Eq. (\ref{EqNoiseSpec}) in Eq. (\ref{EqFGR}), we can write
\begin{eqnarray}
    \Gamma_{g \rightarrow e}^{\rm cap}
    =  2 \left(R_{Q}{\rm Re}[Y [\omega_{ge}]] \right) |\langle e|\varphi|g\rangle|^{2}  \omega_{ge},
\label{EqGammaPhi}
\end{eqnarray}
where $R_{Q} = \hbar/(2e)^{2} \approx 1\; {\rm k}\Omega$ is the \emph{superconducting resistance quantum}.
\par
Since a Josephson junction in the large $E_{J}/E_{C}$ regime (transmon regime) can be modelled as a weakly anharmonic oscillator, there is an easy and straightforward method to evaluate the matrix elements of a transmon-like system using arguments pertinent for a harmonic oscillator. In the limit of weak anharmonicity ($E_{C}/\omega_{ge} \ll1$), the nonlinearity of the transmon leads to only small corrections that can be ignored for our purpose. Hence,
\begin{eqnarray}
    |\langle e | \Phi_{q} | g \rangle | ^{2}& =& \langle g | \Phi_{q} |e \rangle  \langle e | \Phi_{q} | g \rangle \nonumber\\
    &= & \sum_{j} \langle g | \Phi_{q} |j \rangle  \langle j | \Phi_{q} | g \rangle \nonumber\\
    & =&  \langle g |\ \Phi_{q}^{2} | g \rangle = (\Phi_{q}^{ZPF})^{2},
\label{EqHOmat}
\end{eqnarray}
where $\Phi_{q}^{ZPF}$ denotes the zero point flux-fluctuations of the transmon oscillator.
\par
It is straightforward to generalize the matrix element calculation, as outlined above, to higher energy levels using again the harmonic oscillator intuition. This relies on using magnitude of zero point fluctuations as a natural scale to define flux and charge operators for the JJ oscillator just as position and momentum operators for a harmonic oscillator, i.e.,
\begin{eqnarray}
	\Phi_{q} =\Phi_{q}^{ZPF} (a + a^{\dagger}).
\end{eqnarray}
Using this expression, one can easily write down the flux matrix element between levels $n$ and $n+1$ as
\begin{eqnarray}
	\langle j+1 | \varphi | j \rangle &=& \frac{1}{\varphi_{0}}\langle j+1 | \Phi_{q} | j \rangle \nonumber\\
& =& \sqrt{j +1} \frac{\Phi_{q}^{ZPF}}{\varphi_{0}} \nonumber\\
&=& \sqrt{\frac{j+1}{2}} \left(\frac{8 E_{C}}{E_{J}}\right)^{1/4}.
\label{EqPhimat}
\end{eqnarray}
Similarly, using $Q_{q} = -i Q_{q}^{ZPF} (a - a^{\dagger})$, the charge matrix element between the transmon levels $n$ and $n+1$ can be written as
\begin{eqnarray}
	\langle j+1 | n | j \rangle &=& \frac{1}{2e}\langle j+1 | Q_{q} | j \rangle \nonumber\\
&=& \sqrt{\frac{j+1}{2}} \left(\frac{E_{J}}{8 E_{C}}\right)^{1/4}.
\label{EqQimat}
\end{eqnarray}
\par
Using Eq. (\ref{EqPhimat}) in Eq. (\ref{EqGammaPhi}), we obtain Eq. (1) of the main text
\begin{eqnarray}
    \Gamma_{g \rightarrow e}^{\rm cap}
    = p_{S} \frac{\omega_{ge}}{Q_{S}} + p_{J}\frac{\omega_{ge}}{Q_{J}},
\label{EqDiel}
\end{eqnarray}
where we have used $E_{C} = e^{2}/(2 C_{\rm tot})$ with $C_{\rm tot} = C_{J} + C_{S}$. The contributions of shunt capacitance $C_{S}$ and junction capacitance $C_{J}$ have been separated, using the parametrization [c.f. Eq. (\ref{EqReYCap})],
\begin{eqnarray}
    {\rm Re}[Y[\omega_{ge}]] = \frac{\omega_{ge} C_{S}}{Q_{S}} + \frac{\omega_{ge} C_{J}}{Q_{J}},
\end{eqnarray}
and corresponding participation ratios
\begin{eqnarray}
    p_{S} = \frac{C_{S}}{C_{\rm tot}}; \quad p_{J} = \frac{C_{J}}{C_{\rm tot}}.
\end{eqnarray}
Note that here we have assumed $C_{S}, C_{J}$ and $Q_{S}, Q_{J}$ to be frequency independent -- an assumption valid for most materials. For typical samples, $C_{S} \sim 60-70$ fF and $C_{J} \sim 3-5$ fF. Thus, $p_{S} \gg p_{J}$ and most of the dielectric loss is entirely accounted for by the external shunt capacitance. A more accurate estimation of participation ratios may be performed using electromagnetic simulation softwares such as HFSS.
%
\subsection{Purcell loss}
%
The Purcell effect is the spontaneous emission of photons from the qubit through the resonator. The resultant decay rate, taking into account only the fundamental mode of the resonator, is given as $\Gamma^{\rm Purcell} = \kappa g^{2}/\Delta^{2}$ \cite{Houck2008}. Here, $\kappa$ denotes the resonator linewidth, $g$ denotes the qubit-resonator coupling and $\Delta = \omega_{ge} -\omega_{\rm res}$ denotes the qubit-resonator detuning. All parameters required for calculating $\Gamma^{\rm Purcell}$  are measured experimentally and are tabulated in Table \ref{Paramtable}.
%
\section{Quality factor of witness resonators}
%
\begin{figure}[t!]
\centering
  \includegraphics[width=0.8\columnwidth]{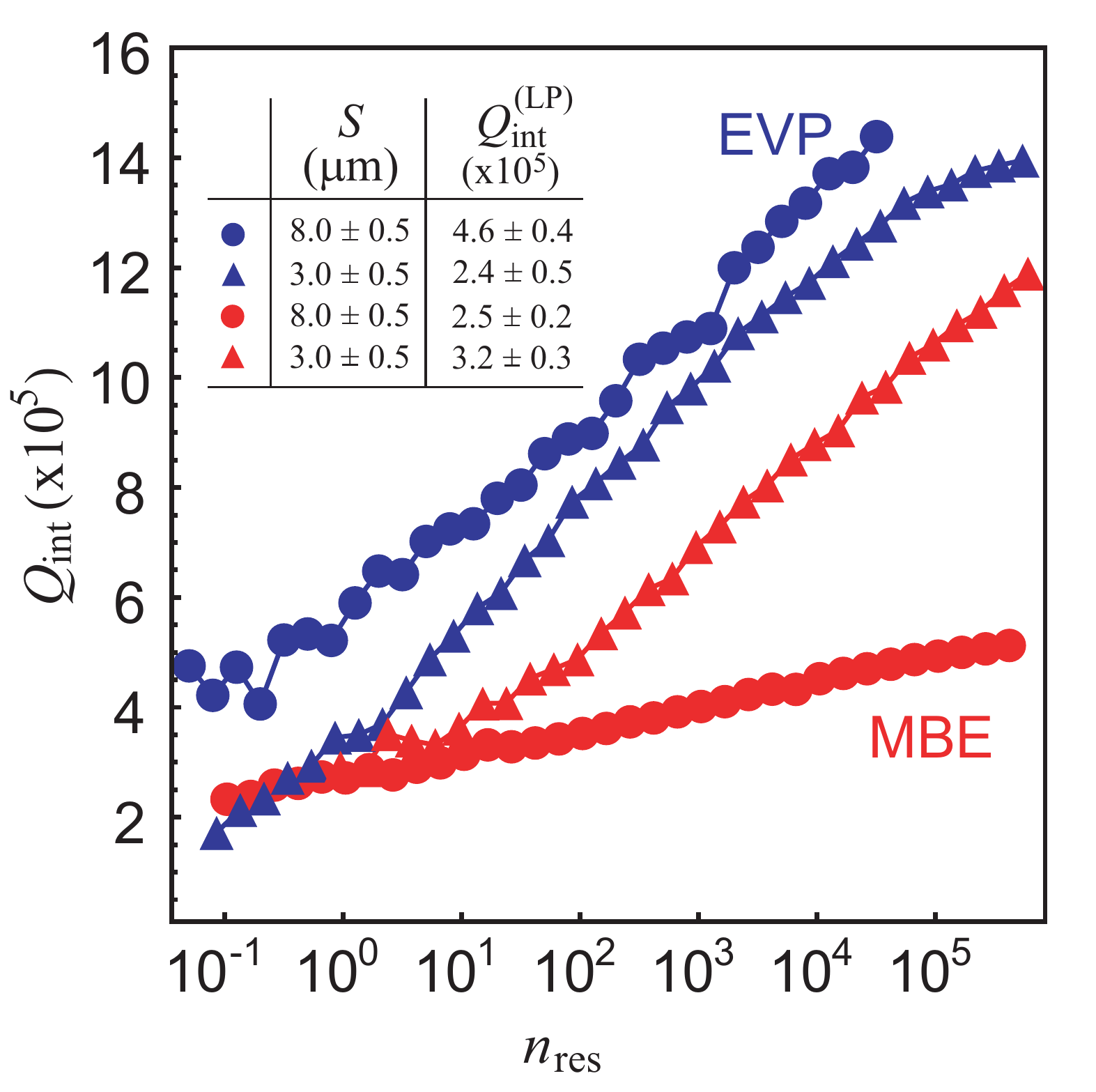}\\
  \caption{{\bf Quality factors of witness resonators measured as  a function of drive power.} The measurements were done on similar chips, each comprising CPW hangers coupled to a feedline. The internal quality factors $Q_{\rm int}$ for both samples decrease with decrease in power (calibrated in terms of equivalent number of photons $n_{\rm res}$), consistent with unsaturated TLS-limited loss at low powers \cite{SageCPW2011}. The inset tabulates the gap $S$ between the ground and center trace of the CPW resonators, along with their mean $Q_{\rm int}$. The standard deviations in $Q_{\rm int}$ are obtained by averaging the five lowest power data points for each set.}
\label{FigRes}
\end{figure}
Measurement of quality factors on witness resonators, fabricated on the same wafers as transmon samples 4,6 (aluminum deposited in higher-vacuum MBE chamber, on annealed sapphire) and 3,5,7 (aluminum deposited in lower-vacuum conventional evaporator, on annealed sapphire), are compiled in Fig. \ref{FigRes}. The quality factors inferred from transmission measurements on CPW hanger resonators \cite{Megrant2012} were found to be in the range of $(2-4)\times 10^{5}$ at resonator powers corresponding to single photon at the resonator frequency. Since, for aluminum devices at low temperatures, the dominant loss can be identified to be capacitive in nature \cite{Vissers2012}, it allows us to do a consistency check of our results by doing a direct comparison of low-power resonator quality factors with those inferred from the dielectric-loss limited transmon $T_{1}$. The difference between low power $Q_{\rm int}$ for witness CPW resonators and $Q_{\rm diel} \geq10^{6}$ for transmons is in accordance with difference between relative surface participation ratios. It can be attributed to the difference in the gap size $S$ between the center trace and the ground plane in CPW resonators ($3-8$ $\mu$m), and the gap between the shunt capacitor fingers ($20-25$ $\mu$m) in the transmon respectively. The role of surface participation has been highlighted in detailed studies performed exclusively on CPW resonators \cite{Pappas2010,Wenner2011, Megrant2012,Quintana2014} previously. 
%
%
\section{Additional AFM and STEM images}
%
%
\begin{figure}[h!]
\centering
  \includegraphics[width=\columnwidth]{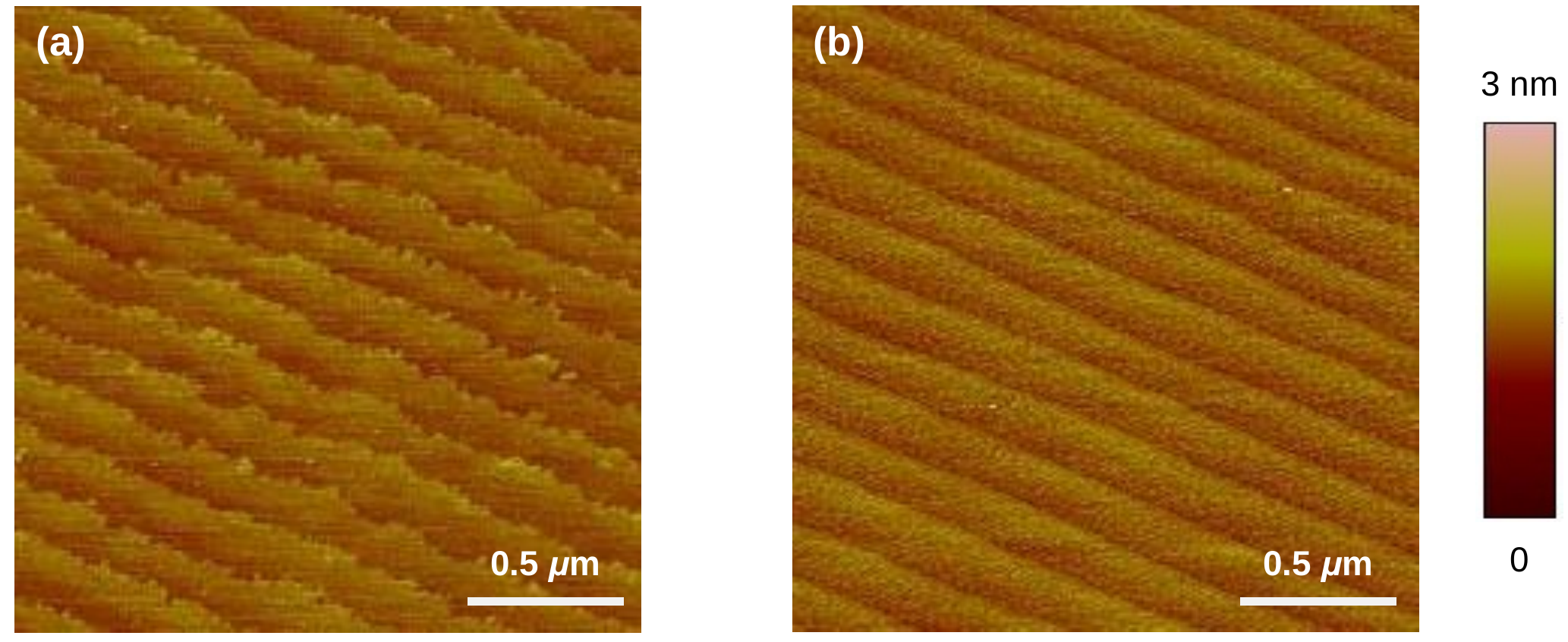}\\
  \caption{{\bf AFM images of sapphire wafer.} Atomic force microscopic images of sapphire wafers (a) pre-annealing (as received from the vendor) and (b) post furnace annealing.}
\label{FigAFM}
\end{figure}
Figure \ref{FigAFM} shows images from atomic force microscopy performed on a typical sapphire wafer, before and after high-temperature anneal. The annealing was performed at 1100~$^{\circ}$C in an atmosphere of ${\rm N_{2}:O_{2}}$, $4:1$. The annealed wafer was then cleaned in a $1:1$ solution of ${\rm H_{2}O_{2}:H_{2}SO_{4}}$. As evident from the images the annealing reduces the surface roughness of sapphire, potentially resulting in better interface quality post-metal deposition.
\par
\begin{figure*}[t!]
\centering
  \includegraphics[width=0.9\textwidth]{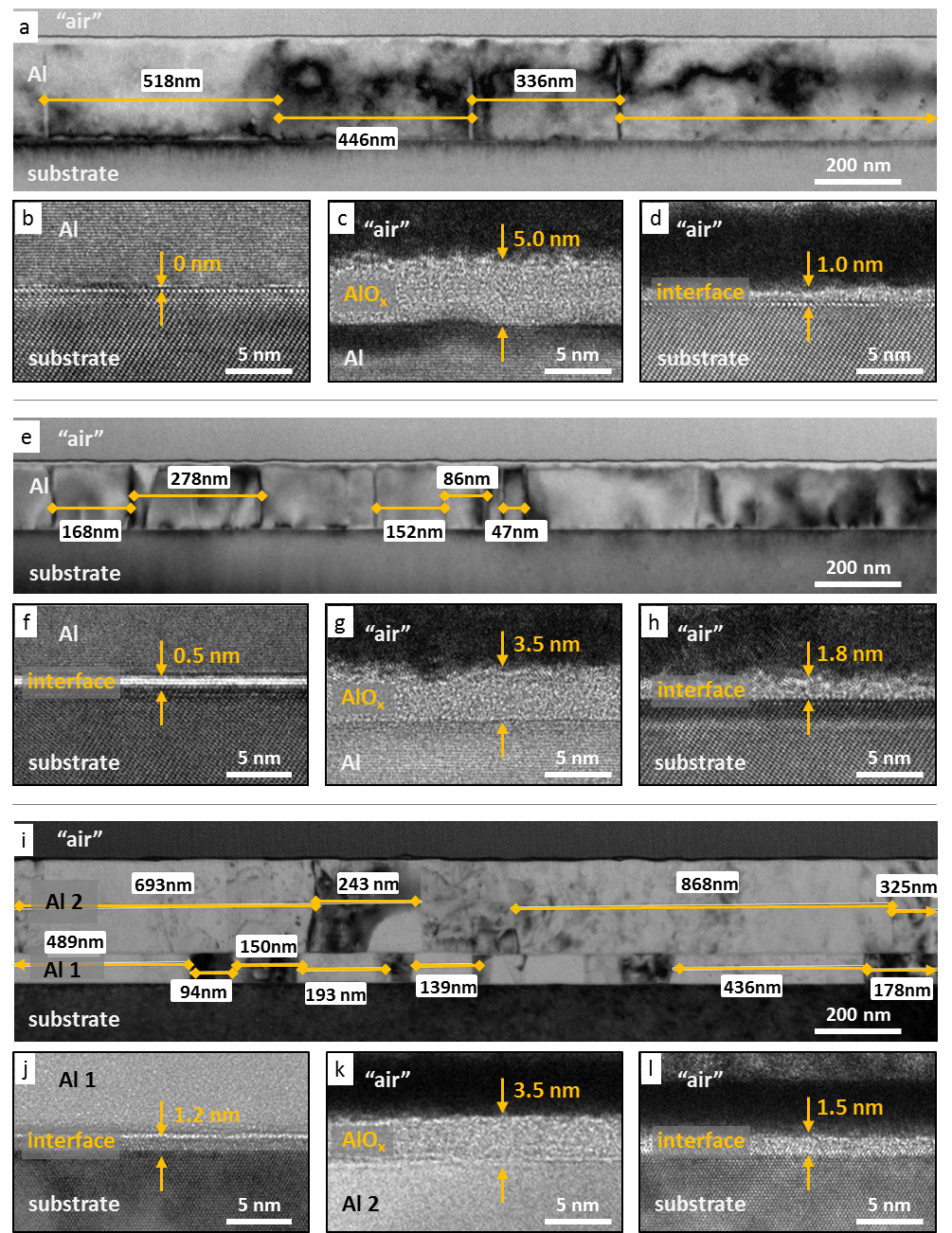}\\
  \caption{{\bf Detailed STEM images of interfaces.} STEM cross-sectional images of (a-d) an MBE sample co-fabricated with sample 6 in the manuscript, (e-h) an evaporated sample co-fabricated with sample 7, and (i-l) an evaporated sample fabricated in the same run and in an identical manner as samples 1 and 2. (a, e, i): Detailed grain structure of each aluminum film.  (b,f,j): High-magnification views of the aluminum--substrate interface. (c,g,k): High-magnification views of the aluminum--``air" interface.  (d,h,l): High-magnification views of the substrate--``air" interface. See the text for additional information.}
\label{FigSTEM}
\end{figure*}

To complement the $T_{1}$ results obtained from qubit samples employed for annealing and deposition test described in the main text, we also used scanning transmission electron microscope (STEM) imaging to investigate the difference in quality of the films and interfaces between these samples. For this purpose, we used the representative samples enlisted below:
\begin{enumerate}
\item MBE annealed: An annealed sapphire substrate with an ultra-high-vacuum (UHV), electron-beam-deposited aluminum (MBE) film that was co-fabricated with sample 6.  Additional sample information is detailed in the manuscript text and in Table 1.
\item Evaporated annealed: An annealed sapphire substrate with a high-vacuum, electron-beam-deposited aluminum (conventional evaporator) film that was co-fabricated with sample 7.  Additional sample information is detailed in the manuscript text and in Table 1.
\item Evaporated non-annealed: A non-annealed sapphire substrate with a double-angle evaporation of high-vacuum, electron-beam-deposited aluminum (conventional evaporator) that was deposited concurrently with Josephson junction growth.  This STEM sample was deposited on a wafer from the same fabrication run, and prepared in an identical manner, as samples 1 and 2.
\end{enumerate}  
Figure 2 in the main text presents images of both annealed samples and included information on the aluminum grains and the aluminum-substrate interfaces. Figure \ref{FigSTEM} includes wider-area images of the aluminum grains for samples presented in the main text, as well as additional high-resolution views of the aluminum--sapphire, aluminum--air, and sapphire--air interfaces for each of the three STEM samples described above.  As a note, prior to preparation of the samples using focused ion beam cross sectioning, the top surface of each sample was coated with a thin layer of iridium (which appears as a dark thin film in the STEM images) as well as a thicker layer of protective platinum. Electron-energy loss spectroscopy (EELS) linescans also were collected for carbon, aluminum, and oxygen peaks across many of the interfaces. EELS linescans are not shown in this supplemental text; however, we summarize the main results of these studies here.
\par
The wide-area images of the aluminum grains for each sample in Figs. \ref{FigSTEM}(a,e,i) indicate that the MBE aluminum grains are larger than the evaporated aluminum grains, which is in line with our expectations based on the deposition rate and method.  
The substrate--aluminum interface for each sample is the critical interface of interest for the annealing test described in the main text.  A key finding was that annealing the substrate provided a substantial improvement in coherence; further aluminum films grown on annealed substrates, irrespective of whether the aluminum was grown in a traditional evaporation chamber or a UHV MBE chamber, resulted in comparable qubit coherence. Figs. \ref{FigSTEM}(b, f, j) detail the interfacial film at the substrate--aluminum interface for the MBE annealed sample (no interface), the evaporated annealed sample (0.5 nm thick interface), and evaporated non-annealed sample (1.2 nm thick interface), respectively. The sharp interface for the MBE annealed sample is exactly as expected: a sharp transition from aluminum + oxygen in the sapphire substrate to aluminum in the MBE aluminum film, as confirmed by the EELS profile study along this interface. For both evaporated samples (annealed and non-annealed), the EELS data indicates that there again is no carbon present in the interfacial layer of either sample; correspondingly, this indicates that the interfacial layer is not hydrocarbon based. The interfacial layer of each evaporated sample does contain both oxygen and aluminum. From this data, it remains unclear whether the interfacial layer only is comprised of aluminum and oxygen, or whether there would be a non-hydrocarbon-based contribution to the layer. The aluminum--air interfaces for each sample are shown in Figs. \ref{FigSTEM}(c,g,k).  EELS data at each interface indicates the presence of both aluminum and oxygen in the interfacial layer, which is consistent with an expected self-passivating aluminum oxide of approximately 3 nm in thickness. The substrate--air interfacial thicknesses were also comparable in all samples [Figs. \ref{FigSTEM}(d,h,l)].  EELS data collected for the MBE annealed and evaporated annealed samples at this interface showed no presence of carbon, while aluminum and oxygen both were found within the interfacial layer.
%
%
%
%
%
%

%